# Supersymmetric Vector Multiplets in Non-Adjoint Representations of SO(N)

Hitoshi NISHINO[1] and Subhash RAJPOOT[2]

*Department of Physics & Astronomy*
*California State University*
*1250 Bellflower Boulevard*
*Long Beach, CA 90840*

## Abstract

In the conventional formulation of $N=1$ supersymmetry, a vector multiplet is supposed to be in the adjoint representation of a given gauge group. We present a new formulation with a vector multiplet in the *non*-adjoint representation of $SO(N)$ gauge group. Our basic algebra is $[T^I, T^J] = f^{IJK}T^K$, $[T^I, U^i] = -(T^I)^{ij}U^j$, $[U^i, U^j] = -(T^I)^{ij}T^I$, where $T^I$ are the generators of $SO(N)$, while $U^i$ are the new 'generators' in certain non-adjoint real representation $R$ of $SO(N)$. We use here the word 'generator' in the broader sense of the word. Such a representation can be any real representation of $SO(N)$ with the positive definite metric, satisfying $(T^I)^{ij} = -(T^I)^{ji}$ and $(T^I)^{[ij|}(T^I)^{|k]l} \equiv 0$. The first non-trivial examples are the spinorial $\mathbf{8}_S$ and conjugate spinorial $\mathbf{8}_C$ representations of $SO(8)$ consistent with supersymmetry. We further couple the system to chiral multiplets, and show that a Higgs mechanism can give positive definite $(\text{mass})^2$ to the new gauge fields for $U^i$. We show an analogous system working with $N=1$ supersymmetry in 10D, and thereby $N=4$ system in 4D interacting with extra multiplets in the representation $R$. We also perform superspace reformulation as an independent confirmation.

PACS: 11.30.Pb, 12.60.Jv, 11.15.-q, 14.70.-e, 11.10.Kk

Key Words: $N=1$ and $N=4$ Supersymmetry, Yang-Mills, Vector Multiplets, Non-Adjoint Vector Fields, Higgs Mechanism, Four and Ten Dimensions.

[1] E-Mail: hnishino@csulb.edu
[2] E-Mail: rajpoot@csulb.edu

## 1. Introduction

In $N=1$ supersymmetric theories in four dimensions (4D), common wisdom tells us that a vector multiplet (VM) [1] has to be in the adjoint representation of a given gauge group. This is the so-called Yang-Mills multiplet, when the gauge group is non-Abelian, dedicated to the initiators of non-Abelian vector fields in physics [2]. Even without supersymmetry, the common practice dictates that a vector field should always be in the adjoint representation, when the group is non-Abelian. Also differential geometrical concepts imply that a gauge group is a manifold, where the connection 1-form $A$ yields the 'curvature' two form $F$ through the relationship $F \equiv dA + A \wedge A$ [3]. By definition, such a 1-form field is a Yang-Mills 'vector' field. However, this does *not* necessarily mean the *non*-existence of a vector field in the *non*-adjoint representation in general.

In this paper, we take the first step to establish $N=1$ supersymmetric VM in the non-adjoint representation of the arbitrary $SO(N)$ gauge group. Our basic algebra has the new generators $U^i$ belonging to a real representation $R$ of $SO(N)$, satisfying certain matrix conditions with the usual $SO(N)$ generators $T^I$. We show that the vectorial representation is the simplest example which, however, has been kind of known since 1970's in the context of 'hidden symmetries'. As the first non-trivial examples, we show that the spinorial representations of $SO(8)$ gauge group satisfy the required conditions. We couple these VMs to chiral multiplets (CMs) [1] in the adjoint and the real representation $R$, and show that the new gauge field for $U^i$ can get masses *via* the Higgs mechanism. We show that a similar system can be formulated in 10D. We also perform a superspace reformulation of the results in section 2.

## 2. Lagrangian for VM in the $\underset{\sim}{\mathbf{N}}$ of SO(N)

There are two basic VMs in our system, the usual VM $(A_\mu{}^I, \lambda^I; D^I)$ and the new VM $(B_\mu{}^i, \chi^i; H^i)$ in a certain appropriate *real* representation $R$ of $SO(N)$. Here the indices $I, J, \cdots = 1, 2, \cdots, N(N-1)/2$ are for the adjoint representation of $SO(N)$, while $i, j, \cdots = 1, 2, \cdots, \dim R$ are for the *real* representation $R$ of $SO(N)$ with a positive definite metric. We do not specify the representation $R$ at this stage, but the simplest example is the vectorial $\mathbf{N}$ representation of $SO(N)$. In such a case, $i, j, \cdots = 1, 2, \cdots, N$. We use always the superscripts for these indices, because the metric is positive definite, and there is no need to distinguish raising or lowering of these indices.

These fields are the field representation of our new algebra satisfying the commutators

$$[T^I, T^J] = +f^{IJK}T^K \ , \tag{2.1a}$$

$$[T^I, U^j] = -(T^I)^{jk}U^k \ , \tag{2.1b}$$

$$[U^i, U^j] = -(T^K)^{ij}T^K \ . \tag{2.1c}$$

Even though common wisdom dictates that all the generators are in adjoint representations, we use the term 'generators' in the sense that (2.1) satisfies Jacobi identities. As has been mentioned, the simplest example is the vectorial **N** representation of $SO(N)$. For this vectorial representation, (2.1) can be rewritten as $[T^{ij}, T^{kl}] = 2\delta^{k[j}T^{i]l} - (k\leftrightarrow l)$, $[T^{ij}, U^k] = \delta^{jk}U^i - \delta^{ik}U^j$ and $[U^i, U^j] = -T^{ij}$. However, this example is kind of trivial, because this is nothing but expressing the algebra of $SO(N+1)$ in terms of $SO(N)$-explicit, but $SO(N+1)$-implicit notation. This can be seen as follows: Let $T^{\hat{i}\hat{j}}$ be the generators of $SO(N+1)$. Among the indices $\hat{i}, \hat{j}, \cdots = 1, 2, \cdots, N+1$, we separate the $(N+1)$-th one, and use the indices $i, j, \cdots = 1, 2, \cdots, N$ for the rest. By identifying the generators $U^i \equiv T^{i,N+1}$, we can re-express the original $SO(N+1)$ commutators, yielding exactly the same commutators as above in terms of $T^{ij}$ and $U^i$, which are manifest in $SO(N)$, but *not* in $SO(N+1)$. In other words, (2.1) is nothing but $SO(N+1)$ algebra, when the indices $i, j, \cdots$ are for the vectorial representations of $SO(N)$.

In fact, this has been known in supergravity since 1970's as 'hidden' symmetries. For example in $N = 7$ supergravity, there are vector fields in the adjoint **21** representation, and in the vectorial **7** representation of $SO(7)$. However, this $N = 7$ supergravity has actually 'hidden' $SO(8)$ symmetry, and the whole system is promoted to $N = 8$ supergravity, where the total $21+7 = 28$ vectors now belong to the adjoint representation of the promoted gauge group $SO(8)$. Another example is adding the spinorial **128** representation of $SO(16)$ to its adjoint **120** representation, forming in total the adjoint **248** representation of a larger group $E_8$. As these examples of the enlarged groups show, we are effectively dealing with groups larger than $SO(N)$.

We mention another important aspect of our system. Due to the algebra (2.1c), once the generators $U^i$ have local parameters with its own gauge field $B_\mu{}^i$, then the generators $T^I$ should be also local with its own gauge fields $A_\mu{}^I$. To be more specific, algebra (2.1c) implies that when the parameters $\beta^i$ for $U^i$ are $x$-dependent, the parameters $\alpha^I$ must be also $x$-dependent. In other words, we can not dispense with the ordinary gauge fields $A_\mu{}^I$, once we introduce the gauge fields $B_\mu{}^i$. As such, we have to maintain the usual VM $(A_\mu{}^I, \lambda^I; D^I)$, once we consider the local symmetry $\delta_U$ with the new VM $(B_\mu{}^i, \chi^i; H^i)$.

Keeping these points in mind, we first present the main results, *i.e.*, the lagrangian of our action $I_{\rm VM} \equiv \int d^4x\, \mathcal{L}_{\rm VM}$ with

$$\mathcal{L}_{\rm VM} = -\tfrac{1}{4}(\mathcal{F}_{\mu\nu}{}^I)^2 + \tfrac{1}{2}(\overline{\lambda}^I \slashed{\mathcal{D}} \lambda^I) + \tfrac{1}{2}(D^I)^2 - \tfrac{1}{4}(G_{\mu\nu}{}^i)^2 + \tfrac{1}{2}(\overline{\chi}^i \slashed{\mathcal{D}} \chi^i) + \tfrac{1}{2}(H^i)^2 \ . \tag{2.2}$$

Even though this lagrangian formally looks the same as that of conventional VMs, the field strengths and covariant derivatives are defined by

$$\mathcal{F}_{\mu\nu}{}^I \equiv +F_{\mu\nu}{}^I - g(T^I)^{ij} B_\mu{}^i B_\nu{}^j \equiv + \left[2\partial_{[\mu} A_{\nu]}{}^I + g f^{IJK} A_\mu{}^J A_\nu{}^K \right] - g(T^I)^{ij} B_\mu{}^i B_\nu{}^j \ , \tag{2.3a}$$

$$G_{\mu\nu}{}^i \equiv +D_\mu B_\nu{}^i - D_\nu B_\mu{}^i \equiv +2\partial_{[\mu} B_{\nu]}{}^i + 2g(T^I)^{ij} A_{[\mu}{}^I B_{\nu]}{}^j \ , \tag{2.3b}$$

$$\mathcal{D}_\mu \chi^i \equiv +D_\mu \chi^i - g(T^I)^{ij} B_\mu{}^j \lambda^I \equiv + \left[\partial_\mu \chi^i + g(T^I)^{ij} A_\mu{}^I \chi^j \right] - g(T^I)^{ij} B_\mu{}^j \lambda^I \ , \tag{2.3c}$$

$$\mathcal{D}_\mu \lambda^I \equiv +D_\mu \lambda^I - g(T^I)^{ij} B_\mu{}^i \chi^j \equiv + \left[\partial_\mu \lambda^I + g f^{IJK} A_\mu{}^J \lambda^K \right] - g(T^I)^{ij} B_\mu{}^i \chi^j \ . \tag{2.3d}$$

The $g$ is the minimal gauge coupling constant, $D_\mu$'s is the usual $SO(N)$ covariant derivative, $F_{\mu\nu}{}^I$ is the usual $SO(N)$ field strength, while $\mathcal{F}_{\mu\nu}{}^I$ and $\mathcal{D}_\mu$ are the fully covariant both under $T^I$ and $U^i$.

Our action $I_{\rm VM}$ is invariant under all the symmetries in the system, $N=1$ supersymmetry $\delta_Q$, $SO(N)$ symmetry $\delta_T$, and new $\delta_U$ symmetry. The first of these has the transformation rule

$$\delta_Q A_\mu{}^I = +(\overline{\epsilon} \gamma_\mu \lambda^I) \ , \tag{2.4a}$$

$$\delta_Q \lambda^I = +\tfrac{1}{2}(\gamma^{\mu\nu} \epsilon) \mathcal{F}_{\mu\nu}{}^I - i(\gamma_5 \epsilon) D^I \ , \tag{2.4b}$$

$$\delta_Q D^I = +i(\overline{\epsilon} \gamma_5 \slashed{\mathcal{D}} \lambda^I) \ , \tag{2.4c}$$

$$\delta_Q B_\mu{}^i = +(\overline{\epsilon} \gamma_\mu \chi^i) \ , \tag{2.4d}$$

$$\delta_Q \chi^i = +\tfrac{1}{2}(\gamma^{\mu\nu} \epsilon) G_{\mu\nu}{}^i - i(\gamma_5 \epsilon) H^i \ , \tag{2.4e}$$

$$\delta_Q H^i = +i(\overline{\epsilon} \gamma_5 \slashed{\mathcal{D}} \chi^i) \ . \tag{2.4f}$$

Note that these transformation rules look formally the same as the conventional rules for VMs, *except* the involvement of the new field strengths $\mathcal{F}$, $G$ and covariant derivative $\mathcal{D}$.

Our action $I_{\rm VM}$ is also invariant under the usual $SO(N)$ transformation $\delta_T$ with the infinitesimal parameter $\alpha^I$:

$$\delta_T A_\mu{}^I = D_\mu \alpha^I \equiv \partial_\mu \alpha^I + g f^{IJK} A_\mu{}^J \alpha^K \ , \tag{2.5a}$$

$$\delta_T \lambda^I = -g f^{IJK} \alpha^J \lambda^K ~, \tag{2.5b}$$

$$\delta_T D^I = -g f^{IJK} \alpha^J D^K ~, \tag{2.5c}$$

$$\delta_T B_\mu{}^i = -g(T^I)^{ij} \alpha^I B_\mu{}^j ~, \tag{2.5d}$$

$$\delta_T \chi^i = -g(T^I)^{ij} \alpha^I \chi^j ~, \tag{2.5e}$$

$$\delta_T H^i = -g(T^I)^{ij} \alpha^I H^j ~, \tag{2.5e}$$

and the new $U^i$-transformation $\delta_U$ with the infinitesimal parameter $\beta^i$:

$$\delta_U A_\mu{}^I = +g(T^I)^{ij} \beta^i B_\mu{}^j ~, \tag{2.6a}$$

$$\delta_U \lambda^I = +g(T^I)^{ij} \beta^i \chi^j ~, \tag{2.6b}$$

$$\delta_U D^I = +g(T^I)^{ij} \beta^i H^j ~, \tag{2.6c}$$

$$\delta_U B_\mu{}^i = +D_\mu \beta^j \equiv \partial_\mu \beta^i + g(T^I)^{ij} A_\mu{}^I \beta^j ~, \tag{2.6d}$$

$$\delta_U \chi^i = +g(T^I)^{ij} \beta^j \lambda^I ~, \tag{2.6e}$$

$$\delta_U H^i = +g(T^I)^{ij} \beta^j D^I ~. \tag{2.6f}$$

As this rule shows, the $\delta_U$-transformation exchanges the two VMs $(A, \lambda; D)$ and $(B, \chi; H)$. Relevantly, a similar property can be found in the $\mathcal{F}$, $G$, $\mathcal{D}\lambda$ and $\mathcal{D}\chi$ transforming under $\delta_U$:

$$\delta_U \mathcal{F}_{\mu\nu}{}^I = +g(T^I)^{ij} \beta^i G_{\mu\nu}{}^j ~, \tag{2.7a}$$

$$\delta_U G_{\mu\nu}{}^i = +g(T^I)^{ij} \beta^j \mathcal{F}_{\mu\nu}{}^I ~, \tag{2.7b}$$

$$\delta_U (\mathcal{D}_\mu \lambda^I) = +g(T^I)^{ij} \beta^i (\mathcal{D}_\mu \chi^j) ~, \tag{2.7c}$$

$$\delta_U (\mathcal{D}_\mu \chi^i) = +g(T^I)^{ij} \beta^j (\mathcal{D}_\mu \lambda^I) ~, \tag{2.7d}$$

The field strengths $\mathcal{F}$ and $G$ also satisfy the Bianchi identities

$$\mathcal{D}_{[\mu} \mathcal{F}_{\nu\rho]}{}^I \equiv D_{[\mu} F_{\nu\rho]}{}^I - g(T^I)^{ij} B_{[\mu}{}^i G_{\nu\rho]}{}^j \equiv 0 ~, \tag{2.8a}$$

$$\mathcal{D}_{[\mu} G_{\nu\rho]}{}^i \equiv D_{[\mu} G_{\nu\rho]}{}^i - g(T^I)^{ij} B_{[\mu}{}^j \mathcal{F}_{\nu\rho]}{}^I \equiv 0 ~, \tag{2.8b}$$

We can confirm the off-shell closure of these algebras, in particular, two supersymmetries close off-shell without any field equations:

$$[\delta_Q(\epsilon_1), \delta_Q(\epsilon_2)] = \delta_P(\xi_3) + \delta_T(\alpha_3) + \delta_U(\beta_3) ~,$$

$$\xi_3^\mu \equiv +2(\bar{\epsilon}_1 \gamma^\mu \epsilon_2) ~, \quad \alpha_3^I \equiv -\xi_3^\mu A_\mu{}^I ~, \quad \beta_3^i \equiv -\xi_3^\mu B_\mu{}^i ~. \tag{2.9}$$

where $\delta_P$ is the usual translation operator.

The supersymmetric action invariance $\delta_Q I_{\text{VM}} = 0$ is confirmed, when the *real* representation $R$ for the indices $i, j, \cdots = 1, 2, \cdots, d \equiv \dim R$ satisfies the three conditions

$$\eta^{ij} = \delta^{ij} \ , \tag{2.10a}$$

$$(T^I)^{ij} = -(T^I)^{ji} \ , \tag{2.10b}$$

$$(T^I)^{[ij|}(T^I)^{|k]l} \equiv 0 \ . \tag{2.10c}$$

Eq. (2.10a) is nothing but the positive definiteness of the metric $\eta^{ij}$ for $R$, while (2.10b) is the antisymmetry of the generator matrices, and (2.10c) is the most crucial for the action invariance $\delta_Q I_{\text{VM}} = 0$. Note that (2.10c) is also equivalent to $(T^I)^{[ij|}(T^I)^{|kl]} \equiv 0$.

We can analyze the condition (2.10c) in terms of group theoretical language. If we introduce the symbols $d \equiv \dim R$ and $I_2(R)$ for the dimensionality and the second index for the representation $R$ normalized as [4]

$$(T^I T^I)^{ij} = -2I_2(R)\delta^{ij} \ , \tag{2.11}$$

we get accordingly

$$(T^I T^J)^{ii} = -\frac{4dI_2(R)}{N(N-1)}\delta^{IJ} \ . \tag{2.12}$$

Using these two equations after multiplying (2.10c) by $(T^J)^{jk}$, we see that a necessary condition of (2.10c) is

$$\frac{2dI_2(R)}{N(N-1)} - 2I_2(R) + N - 2 = 0 \ . \tag{2.13}$$

As has been mentioned, the simplest example for $R$ is the $\mathbf{N}$ representation of $SO(N)$, satisfying (2.13) by $d = \dim(\mathbf{N}) = N$ and $I_2(\mathbf{N}) = (N-1)/2$ [4]. However, there are other non-trivial representations, as well. We show that such non-trivial examples are the spinorial representation $\mathbf{8}_S$ and the conjugate spinorial representation $\mathbf{8}_C$ of $SO(8)$. This is because the metric tensor for the spinorial representation is positive definite [5], and the $SO(8)$ generators in these representations are antisymmetric. Finally, we see that they satisfy the crucial condition (2.10c) for two reasons. First, $I_2(\mathbf{8}_S) = 7/2$ [4] and $d = \dim(\mathbf{8}_S) = 8$ satisfy (2.13) as a necessary condition. Second, more rigorously, because the spinorial matrix representations for the $SO(8)$ generators are nothing but the $\gamma$-matrices satisfying the Clifford algebra of $SO(8)$. These facts can be confirmed by [5], in particular, we can study the Euclidian case of $D=8+0$, and see that the charge conjugation matrix $C$ is symmetric, while $\gamma$-matrices $\gamma^a$ or its antisymmetric products $\gamma^{abcd}$ $(a, b, \cdots = 1, 2, \cdots, 8)$ are all symmetric, and $\gamma^{ab}$ are antisymmetric [5]. Finally, the satisfaction of (2.10c) is understood as follows. We start with the Fierz identity

$$\delta^{AC}\delta^{BD} = +\tfrac{1}{8}\delta^{AB}\delta^{CD} - \tfrac{1}{16}(\gamma^{ab})^{AB}(\gamma^{ab})^{CD} + \tfrac{1}{384}(\gamma^{abcd})^{AB}(\gamma^{abcd})^{CD} \quad , \tag{2.14}$$

with the indices $A, B, \cdots, = 1, 2, \cdots, 8$ for $\mathbf{8}_\mathrm{S}$ instead of $i, j, \cdots$. Now if we take the $[ABC]$ components of both sides of (2.14), only the second term on the r.h.s. remains satisfying (2.10c): $(\gamma^{ab})^{[AB|}(\gamma^{ab})^{|C]D} \equiv 0$. In the case of the conjugate $\mathbf{8}_\mathrm{C}$ of $SO(8)$, we can just flip all the undotted indices in (2.14) into the dotted ones, and again (2.10c) is satisfied. Therefore all the conditions in (2.10) are satisfied both for the $\mathbf{8}_\mathrm{S}$ and $\mathbf{8}_\mathrm{C}$ of $SO(8)$.

## 3. Couplings to CMs

After establishing the invariant action under all the required symmetries, the next natural question is whether there is a mechanism of giving the masses to the new gauge field $B_\mu{}^i$. This is because massless gauge fields are not quite acceptable as phenomenological applications. In this section, we do not specify the representation $R$ for the indices $i, j, \cdots$, and do not restrict them to be the vectorial representation of $SO(N)$, even though the latter is the simplest example for illustrative purposes.

To this end, we couple our basic action $I_\mathrm{VM}$ to CMs. The important point is that such new interactions should be also invariant under the $\delta_U$-transformations. The natural choice is the CMs both in the adjoint and vectorial representations, i.e. $(A^I, B^I, \psi^I; F^I, G^I)$ and $(A^i, B^i, \psi^i; F^i, G^i)$. This is because the $\delta_U$-transformation exchanges these multiplets. In order to write down the cubic interactions, however, we need an additional extra singlet CM $(A, B, \psi; F, G)$ neutral both under $\delta_T$ and $\delta_U$.

The action for the kinetic terms for these three CMs is $I_\mathrm{CM} \equiv \int d^4x\, \mathcal{L}_\mathrm{CM}$, where

$$\begin{aligned}\mathcal{L}_\mathrm{CM} =\ & -\tfrac{1}{2}(\mathcal{D}_\mu A^I)^2 - \tfrac{1}{2}(\mathcal{D}_\mu B^I)^2 + \tfrac{1}{2}(\overline{\psi}^I \slashed{\mathcal{D}} \psi^i) + \tfrac{1}{2}(F^I)^2 + \tfrac{1}{2}(G^I)^2 \\ & -\tfrac{1}{2}(\mathcal{D}_\mu A^i)^2 - \tfrac{1}{2}(\mathcal{D}_\mu B^i)^2 + \tfrac{1}{2}(\overline{\psi}^i \slashed{\mathcal{D}} \psi^i) + \tfrac{1}{2}(F^i)^2 + \tfrac{1}{2}(G^i)^2 \\ & -\tfrac{1}{2}(\partial_\mu A)^2 - \tfrac{1}{2}(\partial_\mu B)^2 + \tfrac{1}{2}(\overline{\psi} \slashed{\partial} \psi) + \tfrac{1}{2}F^2 + \tfrac{1}{2}G^2 \\ & - g f^{IJK}(\overline{\lambda}^I \psi^J) A^K - i g f^{IJK}(\overline{\lambda}^I \gamma_5 \psi^J) B^K + (T^I)^{ij}(\overline{\lambda}^I \psi^i) A^j + i g (T^I)^{ij}(\overline{\lambda}^I \gamma_5 \psi^i) B^j \\ & + g f^{IJK} D^I A^J B^K - g(T^I)^{ij} D^I A^i B^j - g(T^I)^{ij}(\overline{\psi}^I \chi^i) A^j - i g (T^I)^{ij}(\overline{\psi}^I \gamma_5 \chi^i) B^j \\ & - g(T^I)^{ij}(\overline{\psi}^i \chi^j) A^I - i g (T^I)^{ij}(\overline{\psi}^i \gamma_5 \chi^j) B^I + g(T^I)^{ij} H^i (A^I B^j - A^j B^I) \, . \end{aligned} \tag{3.1}$$

The covariant derivatives are defined by

$$\mathcal{D}_\mu \Phi^I \equiv D_\mu \Phi^I - g(T^I)^{ij} B_\mu{}^i \Phi^j \quad, \quad \mathcal{D}_\mu \Phi^i \equiv D_\mu \Phi^i - g(T^I)^{ij} B_\mu{}^j \Phi^I \quad, \tag{3.2}$$

where $\Phi^I$ and $\Phi^i$ represent any component fields in $(A^I, B^I, \chi^I; F^I, G^I)$ and $(A^i, B^i, \chi^i; F^i, G^i)$, respectively. Our action $I_{\text{CM}}$ is invariant under supersymmetry

$$\delta_Q A^A = +(\bar{\epsilon}\psi^A) \quad, \quad \delta_Q B^A = +i(\bar{\epsilon}\gamma_5\psi^A) \quad, \tag{3.3a}$$

$$\delta_Q \psi^A = -(\gamma^\mu \epsilon)\mathcal{D}_\mu A^A + i(\gamma_5\gamma^\mu\epsilon)\mathcal{D}_\mu B^A - \epsilon F^A - i(\gamma_5\epsilon)G^A \quad, \tag{3.3b}$$

$$\delta_Q F^I = +(\bar{\epsilon}\slashed{\mathcal{D}}\psi^I) + gf^{IJK}(\bar{\epsilon}\lambda^J)A^K + igf^{IJK}(\bar{\epsilon}\gamma_5\lambda^J)B^K$$
$$- g(T^I)^{ij}(\bar{\epsilon}\chi^i)A^j - ig(T^I)^{ij}(\bar{\epsilon}\gamma_5\chi^i)B^j \quad, \tag{3.3c}$$

$$\delta_Q G^I = +i(\bar{\epsilon}\gamma_5\slashed{\mathcal{D}}\psi^I) - gf^{IJK}(\bar{\epsilon}\lambda^J)B^K + igf^{IJK}(\bar{\epsilon}\gamma_5\lambda^J)A^K$$
$$+ g(T^I)^{ij}(\bar{\epsilon}\chi^i)B^j - ig(T^I)^{ij}(\bar{\epsilon}\gamma_5\chi^i)A^j \quad, \tag{3.3d}$$

$$\delta_Q F^i = +(\bar{\epsilon}\slashed{\mathcal{D}}\psi^i) + g(T^I)^{ij}(\bar{\epsilon}\lambda^I)A^j + ig(T^I)^{ij}(\bar{\epsilon}\gamma_5\lambda^J)B^j$$
$$- g(T^I)^{ij}(\bar{\epsilon}\chi^j)A^I - ig(T^I)^{ij}(\bar{\epsilon}\gamma_5\chi^j)B^I \quad, \tag{3.3e}$$

$$\delta_Q G^i = +i(\bar{\epsilon}\gamma_5\slashed{\mathcal{D}}\psi^i) - g(T^I)^{ij}(\bar{\epsilon}\lambda^I)B^j + ig(T^I)^{ij}(\bar{\epsilon}\gamma_5\lambda^I)A^j$$
$$+ g(T^I)^{ij}(\bar{\epsilon}\chi^j)B^I - ig(T^I)^{ij}(\bar{\epsilon}\gamma_5\chi^j)A^I \quad, \tag{3.3f}$$

$$\delta_Q F = +(\bar{\epsilon}\slashed{\partial}\psi) \quad, \quad \delta_Q G = +i(\bar{\epsilon}\gamma_5\slashed{\partial}\psi) \quad, \tag{3.3g}$$

where the index $A$ stands for any of the indices $I, i$ or even no index for the multiplet $(A, B, \psi; F, G)$, in order to save space.

Relevantly, we have the supersymmetric mass action $I_m \equiv \int d^4x \, \mathcal{L}_m$ with

$$\mathcal{L}_m \equiv +m\Big[ F^I A^I + F^i A^i + FA + G^I B^I + G^i B^i + GB + \tfrac{1}{2}(\bar{\psi}^I \psi^I) + \tfrac{1}{2}(\bar{\psi}^i \psi^i) + \tfrac{1}{2}(\bar{\psi}\psi) \Big], \tag{3.4}$$

and a typical cubic action $I_{\Phi^3} \equiv \int d^4x \, \mathcal{L}_{\Phi^3}$ with

$$\begin{aligned}\mathcal{L}_{\Phi^3} =\ & +\tfrac{1}{2}\nu F\Big[(A^I)^2 + (A^i)^2 - (B^I)^2 - (B^i)^2\Big] + \nu A(F^I A^I + F^i A^i) - \nu B(F^I B^I + F^i B^i) \\ & + \nu G(A^I B^I + A^i B^i) + \nu A(G^I B^I + G^i B^i) + \nu B(G^I A^I + G^i B^i) \\ & + \tfrac{1}{2}\nu A\Big[(\bar{\psi}^I \psi^I) + (\bar{\psi}^i \psi^i)\Big] - \tfrac{i}{2}\nu B\Big[(\bar{\psi}^I \gamma_5 \psi^I) + (\bar{\psi}^i \gamma_5 \psi^i)\Big] \\ & + \nu\Big[A^I(\bar{\psi}\psi^I) + A^i(\bar{\psi}\psi^i)\Big] - i\nu\Big[B^I(\bar{\psi}\gamma_5\psi^I) + B^i(\bar{\psi}\gamma_5\psi^i)\Big] \quad. \end{aligned} \tag{3.5}$$

The $\nu$'s is a real cubic coupling constant. The form of these cubic couplings is not unique, and is just a simple example. In fact, we could put different coupling constants between these three CMs, and we could also put purely singlet cubic terms of $\Phi^3$ without any $SO(N)$ indices.

All of these actions are invariant under $\delta_Q$, $\delta_T$, and also the $\delta_U$-transformation

$$\delta_U \Phi^I = +g(T^I)^{ij}\beta^i \Phi^j \ , \quad \delta_U \Phi^i = +g(T^I)^{ij}\beta^j \Phi^I \ , \quad \delta_U \Phi = 0 \ , \tag{3.6}$$

where $\Phi$ stands for the multiplet $(A, B, \psi; F, G)$. Similarities of our system to the conventional CM couplings [6] are such as the $g\lambda^I \psi^i A^j$-term, while differences are found in terms with interactions with $B_\mu{}^i$ or $H^i$ in $\mathcal{L}_{\rm CM}$, or any couplings required by the $\delta_U$-invariance.

We mention the issue of uniqueness of the couplings between our two VMs and CMs. As for the number of CMs, it seems that at least two CMs $\Phi^I$ and $\Phi^i$ are needed. This is because the index $i$ on $B_\mu{}^i$ should be contracted in the two equations in (3.2), which are supposed to be covariant under the $\delta_U$-transformation. In this sense, it seems that our lagrangian (3.1) is the minimal form for the kinetic terms with $\Phi^I$ and $\Phi^i$. These kinetic terms and mass terms (3.4) do not require the neutral CM $\Phi$, which is needed for the first time to build the cubic interactions (3.5). This is clear, because there is no way to form an invariant cubic potential action out of two CMs $\Phi^I$ and $\Phi^i$. We can dispense with the neutral $\Phi$, if there is an invariant constant tensor with the index structure $C^{Iij}$, but there seems to be no such a tensor. For example, $(T^I)^{ij}$ can not play such a role, because of the antisymmetry in $i \leftrightarrow j$ yielding the vanishing result for $(T^I)^{ij}\Phi^I\Phi^i\Phi^j \equiv 0$.

## 4. Higgs Mechanism for Masses of New Gauge Fields

We have so far the total action $I_{\rm total} = I_{\rm VM} + I_{\rm CM} + I_m + I_{\Phi^3}$. In order to study a possible Higgs mechanism, we eliminate all the auxiliary fields $D^I$, $D^i$, $F^I$, $F^i$, $F$, $G^I$, $G^i$ and $G$. After this, we get the positive definite potential:

$$\begin{aligned} V = &+ \tfrac{1}{2}\left[mA + \tfrac{1}{2}\nu\{(A^I)^2 + (A^i)^2 - (B^I)^2 - (B^i)^2\}\right]^2 + \tfrac{1}{2}\left[mB + \nu(A^I B^I + A^i B^i)\right]^2 \\ &+ \tfrac{1}{2}\left[(m+\nu A)A^I - \nu B B^I\right]^2 + \tfrac{1}{2}\left[(m+\nu A)B^I + \nu A^I B\right]^2 + \tfrac{1}{2}\left[(m+\nu A)A^i - \nu B B^i\right]^2 \\ &+ \tfrac{1}{2}\left[(m+\nu A)B^i + \nu A^i B\right]^2 + \tfrac{1}{2}g^2\left[f^{IJK}A^J B^K - (T^I)^{ij}A^i B^j\right]^2 \\ &- \tfrac{1}{2}g^2(T^I T^J)^{ij}(A^I B^i - A^i B^I)(A^J B^j - A^j B^J) \ , \end{aligned} \tag{4.1}$$

The negative sign for the last term is due to the antisymmetry of $T^I$, but this term is positive definite as a whole. Note that the representation $R$ for the indices $i, j, \cdots = 1, 2, \cdots, \dim R \equiv d$ has not been specified.

Since the potential (4.1) is positive definite, we can maintain supersymmetry, while breaking the $SO(N)$ symmetry, iff the following eight simultaneous equations are satisfied:

$$\langle A^I\rangle^2 + \langle A^i\rangle^2 = \langle B^I\rangle^2 + \langle B^i\rangle^2 - 2m\nu^{-1}\langle A\rangle \ , \tag{4.2a}$$

$$\langle A^I\rangle\langle B^I\rangle + \langle A^i\rangle\langle B^i\rangle + m\nu^{-1}\langle B\rangle = 0 \ , \tag{4.2b}$$

$$(m + \nu\langle A\rangle)\langle A^I\rangle = \nu\langle B\rangle\langle B^I\rangle \ , \tag{4.2c}$$

$$(m + \nu\langle A\rangle)\langle B^I\rangle + \nu\langle A^I\rangle\langle B\rangle = 0 \ , \tag{4.2d}$$

$$(m + \nu\langle A\rangle)\langle A^i\rangle = \nu\langle B\rangle\langle B^i\rangle \ , \tag{4.2e}$$

$$(m + \nu\langle A\rangle)\langle B^i\rangle + \nu\langle A^i\rangle\langle B\rangle = 0 \ , \tag{4.2f}$$

$$f^{IJK}\langle A^J\rangle\langle B^K\rangle = (T^I)^{ij}\langle A^i\rangle\langle B^j\rangle \ , \tag{4.2g}$$

$$(T^I T^J)^{ij}(\langle A^I\rangle\langle B^i\rangle - \langle A^i\rangle\langle B^I\rangle)(\langle A^J\rangle\langle B^j\rangle - \langle A^j\rangle\langle B^J\rangle) = 0 \ . \tag{4.2h}$$

We next look into the possible non-trivial v.e.v.'s that satisfy all the conditions in (4.2). As the simplest ansatz, we require that

$$\langle B^I\rangle = 0 \ , \quad \langle A^i\rangle = 0 \ , \quad \langle B^i\rangle = 0 \ , \quad \langle B\rangle = 0 \ , \tag{4.3a}$$

$$\langle A\rangle = -m\nu^{-1} \ , \quad \langle A^I\rangle^2 = +2m^2\nu^{-2} \ . \tag{4.3b}$$

These v.e.v.'s easily satisfy all the conditions (4.2a) through (4.2h). This set of solutions is just a simple example, but there may be other sets of more non-trivial solutions.

We next analyze the mass matrices for the vector fields. Here we no longer use the ansatz (4.3), but use general v.e.v.'s. The mass matrices for $A_\mu{}^I$ and $B_\mu{}^i$ can be easily computed by looking into the $(\text{v.e.v.})^2 \times A_\mu^I A^{\mu J}$ or $(\text{v.e.v.})^2 \times B_\mu^i B^{\mu j}$ in the lagrangian $\mathcal{L}_{\text{CM}}$, respectively as

$$(M^2)^{IJ} = 2g^2 h^{IK,JL}(\langle A^K A^L\rangle + \langle B^K B^L\rangle) - 2g^2(T^I T^J)^{ij}(\langle A^i A^j\rangle + \langle B^i B^j\rangle) \ , \tag{4.4a}$$

$$(M^2)^{ij} = 2g^2(T^I)^{ik}(T^I)^{jl}(\langle A^k A^l\rangle + \langle B^k B^l\rangle) - 2g^2(T^I T^J)^{ij}(\langle A^I A^J\rangle + \langle B^I B^J\rangle) \ , \tag{4.4b}$$

where $h^{IJ,KL} \equiv f^{IJM} f^{MKL}$, and $\langle A^K A^L\rangle \equiv \langle A^K\rangle\langle A^L\rangle$, *etc.* to save space. The negative signs for the second terms in (4.4) are due to the antisymmetry of the generators $T^I$.

We can easily confirm that both of these mass matrices have positive definite eigenvalues. We start with $(M^2)^{IJ}$. We first note that $(M^2)^{IJ}$ is rewritten as

$$(M^2)^{IJ} = -2g^2\langle A^K\rangle(T^I T^J)^{KL}\langle A^L\rangle - 2g^2\langle B^K\rangle(T^I T^J)^{KL}\langle B^L\rangle$$
$$- 2g^2\langle A^i\rangle(T^I T^J)^{ij}\langle A^j\rangle - 2g^2\langle B^i\rangle(T^I T^J)^{ij}\langle B^j\rangle = -2g^2 \sum_a \langle a|T^I T^J|a\rangle \ , \tag{4.5}$$

because $f^{IJK} = (T^I)^{JK}$. In the last expression, the bra $\langle a|$ or cket $|a\rangle$ denotes all of the vectors $A^I$, $B^I$, $A^i$ and $B^i$ collectively. Now, since $(M^2)^{IJ}$ is symmetric, it can be diagonalized by orthogonal matrices $\Omega^{IJ}$, satisfying $\Omega^{IK}\Omega^{JK} = \delta^{IJ}$. Let $N^{IJ}$ be the diagonalized mass matrix of $(M^2)^{IJ}$:

$$(M^2)^{IJ} \longrightarrow N^{IJ} = (\Omega M^2 \Omega^T)^{IJ} = \Omega^{IK}(M^2)^{KL}\Omega^{JL} = -2g^2 \sum_a \langle a|\widetilde{T}^I\widetilde{T}^J|a\rangle , \qquad (4.6)$$

where $\widetilde{T}^I \equiv \Omega^{IJ}T^J$. By definition, $N^{IJ}$ has only diagonal components, so that all we have to show is that all the $\not\sum_I N^{II}$ components are positive definite. Here the symbol $\not\sum_I$ implies *no* summation over $I$. In fact, we get the positive definiteness for all $I$, as

$$\begin{aligned}
\sum_I N^{II} &= -2g^2 \sum_I \sum_a \langle a|\widetilde{T}^I\widetilde{T}^I|a\rangle = -2g^2 \sum_I \sum_{a,b} \langle a|\widetilde{T}^I|b\rangle\langle b|\widetilde{T}^I|a\rangle \\
&= +2g^2 \sum_I \sum_{a,b} \langle a|\widetilde{T}^I|b\rangle\langle a|\widetilde{T}^I|b\rangle = +2g^2 \sum_I \sum_{a,b} (\langle a|\widetilde{T}^I|b\rangle)^2 \geq 0 ,
\end{aligned} \qquad (4.7)$$

due to $\langle a|\widetilde{T}^I|b\rangle = -\langle b|\widetilde{T}^I|a\rangle$.

As for $(M^2)^{ij}$, we first rewrite it as

$$(M^2)^{ij} = +2g^2 \left[ \mathcal{A}^{Ii}\mathcal{A}^{Ij} + \mathcal{B}^{Ii}\mathcal{B}^{Ij} - (\mathcal{P}^2)^{ij} - (\mathcal{Q}^2)^{ij} \right] ,$$

$$\mathcal{A}^{Ii} \equiv (T^I)^{ij}\langle A^j\rangle , \quad \mathcal{B}^{Ii} \equiv (T^I)^{ij}\langle B^j\rangle , \quad \mathcal{P}^{ij} \equiv (T^I)^{ij}\langle A^I\rangle , \quad \mathcal{Q}^{ij} \equiv (T^I)^{ij}\langle B^I\rangle . \qquad (4.8)$$

This symmetric matrix $(M^2)^{ij}$ can be diagonalized by an orthogonal matrix $\Lambda^{ij}$, satisfying $\Lambda^{ik}\Lambda^{jk} = \delta^{ij}$:

$$\begin{aligned}
(M^2)^{ij} \longrightarrow N^{ij} &= +(\Lambda M^2 \Lambda^T)^{ij} = \Lambda^{ik}(M^2)^{kl}\Lambda^{jl} \\
&= +2g^2 \Lambda^{ik}\Lambda^{jl} \left[ \mathcal{A}^{Ik}\mathcal{A}^{Il} + \mathcal{B}^{Ik}\mathcal{B}^{Il} - (\mathcal{P}^2 + \mathcal{Q}^2)^{kl} \right] \\
&= +2g^2 \left[ \widetilde{\mathcal{A}}^{Ii}\widetilde{\mathcal{A}}^{Ij} + \widetilde{\mathcal{B}}^{Ii}\widetilde{\mathcal{B}}^{Ij} - (\widetilde{\mathcal{P}}^2 + \widetilde{\mathcal{Q}}^2)^{ij} \right] ,
\end{aligned} \qquad (4.9a)$$

$$\widetilde{\mathcal{A}}^{Ii} \equiv \Lambda^{ij}\mathcal{A}^{Ij} , \quad \widetilde{\mathcal{B}}^{Ii} \equiv \Lambda^{ij}\mathcal{B}^{Ij} , \quad \widetilde{\mathcal{P}}^{ij} \equiv \Lambda^{ik}\Lambda^{jl}\mathcal{P}^{kl} , \quad \widetilde{\mathcal{Q}}^{ij} \equiv \Lambda^{ik}\Lambda^{jl}\mathcal{Q}^{kl} . \qquad (4.9b)$$

In (4.9), use is made of the relation $\Lambda\mathcal{P}^2\Lambda^T = (\Lambda\mathcal{P}\Lambda^T)(\Lambda\mathcal{P}\Lambda^T) = \widetilde{\mathcal{P}}^2$ and *idem.* for $\mathcal{Q}$. Now, what we have to show is that the $\not\sum_i N^{ii}$ are all positive definite. Because of $(\widetilde{\mathcal{P}}^2)^{ij} = -(\widetilde{\mathcal{P}}^2)^{ji}$ and $(\widetilde{\mathcal{Q}}^2)^{ij} = -(\widetilde{\mathcal{Q}}^2)^{ji}$, the $\widetilde{\mathcal{P}}^2$ and $\widetilde{\mathcal{Q}}^2$-terms in $\not\sum_i N^{ii}$ can be rewritten as positive definite square terms. In fact, for all $i$ we get

$$\begin{aligned}
\sum_i N^{ii} &= +2g^2 \sum_i \left[ \widetilde{\mathcal{A}}^{Ii}\widetilde{\mathcal{A}}^{Ii} + \widetilde{\mathcal{B}}^{Ii}\widetilde{\mathcal{B}}^{Ii} + \widetilde{\mathcal{P}}^{ik}\widetilde{\mathcal{P}}^{ik} + \widetilde{\mathcal{Q}}^{ik}\widetilde{\mathcal{Q}}^{ik} \right] \\
&= +2g^2 \sum_i \left[ (\widetilde{\mathcal{A}}^{Ii})^2 + (\widetilde{\mathcal{B}}^{Ii})^2 + (\widetilde{\mathcal{P}}^{ik})^2 + (\widetilde{\mathcal{Q}}^{ik})^2 \right] \geq 0 ,
\end{aligned} \qquad (4.10)$$

This completes the confirmation of the positive definiteness of all the eigenvalues of the mass matrices $(M^2)^{IJ}$ and $(M^2)^{ij}$.

## 5. Applications to 10D and N=4 Supersymmetry in 4D

We have so far dealt with the simple $N=1$ supersymmetry in 4D. We stress, however, that we can apply the same technique to VMs in higher dimensions. The typical example is 10D, where we have the field contents for the on-shell VMs $(A_\mu{}^I, \lambda^I)$ [7] and $(B_\mu{}^i, \chi^i)$, where $\lambda$ and $\chi$ are both Majorana-Weyl spinors of the same chirality in 10D. Our action $I_{10D} \equiv \int d^{10}x \, \mathcal{L}_{10D}$ has the lagrangian formally the same as (2.2), *except for* absent auxiliary fields:

$$\mathcal{L}_{10D} = -\tfrac{1}{4}(\mathcal{F}_{\mu\nu}{}^I)^2 + \tfrac{1}{2}(\overline{\lambda}^I \slashed{\mathcal{D}} \lambda^I) - \tfrac{1}{4}(G_{\mu\nu}{}^i)^2 + \tfrac{1}{2}(\overline{\chi}^i \slashed{\mathcal{D}} \chi^i) \ , \tag{5.1}$$

also with the covariant derivatives and field strengths formally the same as in (2.3). Our action $I_{10D}$ is invariant under formally the same supersymmetry transformation rule as (2.4) *except for* auxiliary fields:

$$\delta_Q A_\mu{}^I = +(\overline{\epsilon}\gamma_\mu \lambda^I) \ , \tag{5.2a}$$

$$\delta_Q \lambda^I = +\tfrac{1}{2}(\gamma^{\mu\nu}\epsilon)\mathcal{F}_{\mu\nu}{}^I \ , \tag{5.2b}$$

$$\delta_Q B_\mu{}^i = +(\overline{\epsilon}\gamma_\mu \chi^i) \ , \tag{5.2c}$$

$$\delta_Q \chi^i = +\tfrac{1}{2}(\gamma^{\mu\nu}\epsilon)G_{\mu\nu}{}^i \ . \tag{5.2d}$$

The reason of no formal difference from 4D is that all the terms arising in $\delta_Q I_{10D}$ cancel exactly in the same way as in 4D, including the Fierz identities

$$(T^I)^{ij}(\overline{\lambda}^I \gamma^\mu \chi^i)(\overline{\epsilon}\gamma_\mu \chi^j) \equiv -\tfrac{1}{2}(T^I)^{ij}(\overline{\epsilon}\gamma_\mu \lambda^I)(\overline{\chi}^i \gamma^\mu \chi^j) \ , \tag{5.3a}$$

$$f^{IJK}(\overline{\epsilon}\gamma_\mu \lambda^I)(\overline{\lambda}^J \gamma^\mu \lambda^K) \equiv 0 \ , \tag{5.3b}$$

which hold both in 4D and 10D. As for the representation $R$, its conditions are the same as in (2.10) for the 4D case.

Due to the absence of auxiliary fields analogous to $D^I$ and $H^i$ in 4D, our system in 10D is an *on-shell* system. It seems that these auxiliary fields are important in 4D for the coupling to chiral multiplets as in (3.1), but not for the covariantized kinetic terms (5.1). A more basic but technical explanation is that for the action invariance of these covariant kinetic terms, the auxiliary fields do not play crucial roles, which is essentially composed of quadratic terms other than the cubic terms that need Fierzings (5.3).

The above result automatically implies that we can have $N=4$ descendant theory in 4D, by simple dimensional reduction [8]. Most importantly, we have extra multiplets in the $\mathbf{8}_\text{S}$ or $\mathbf{8}_\text{C}$ of $SO(8)$ coupled to the *maximal* $N=4$ supersymmetric Yang-Mills multiplet. We stress that this is a surprising feature for such a $N=4$ model, as opposed to the common wisdom that $N=4$ supersymmetric Yang-Mills theory is to be 'maximal', not to be coupled to any multiplets in *non*-adjoint representations.

## 6. Superspace Reformulation

We have so far dealt only with component formulations, so that the next natural step is superspace reformulation [9]. Here we reformulate the 4D result of section 2 in superspace. The local superspace coordinate indices are $A \equiv (a,\alpha)$, $B \equiv (b,\beta)$, $\cdots$, with $a, b, \cdots = 0, 1, 2, 3$ for the bosonic 4D coordinates, and $\alpha, \beta, \cdots = 1, 2, 3, 4$ for the fermionic coordinates. Our basic supercovariant derivative is defined by

$$\nabla_A \equiv D_A + gA_A{}^I T^I + gB_A{}^i U^i ~, \tag{6.1}$$

with $D_A \equiv E_A{}^M \partial_M$ corresponding to '$\mathcal{D}_\mu$' in component formulation. Accordingly, we have the superfield strengths corresponding to $\mathcal{F}_{\mu\nu}{}^I$ and $G_{\mu\nu}{}^i$ in (2.3):

$$[\nabla_A, \nabla_B\} = T_{AB}{}^C \nabla_C + gF_{AB}{}^I T^I + gG_{AB}{}^i U^i ~, \tag{6.2a}$$

$$F_{AB}{}^I \equiv D_{[A} A_{B)}{}^I - T_{AB}{}^C A_C{}^I + gf^{IJK} A_A{}^J A_B{}^K - g(T^I)^{ij} B_A{}^i B_B{}^j ~, \tag{6.2b}$$

$$G_{AB}{}^i \equiv D_{[A} B_{B)}{}^i - T_{AB}{}^C B_C{}^i + g(T^I)^{ij} A_{[A}{}^I B_{B)}{}^j ~. \tag{6.2c}$$

In superspace, we use the antisymmetrization rule, *e.g.*, $M_{[AB)} \equiv M_{AB} - (-1)^{AB} M_{BA}$, *etc.* Accordingly, the Jacobi identity $[\nabla_{[A|}, [\nabla_{|B|}, \nabla_{|C)}\}\} \equiv 0$ yields the Bianchi identities (BIs)

$$\tfrac{1}{2}\nabla_{[A} F_{BC)}{}^I - \tfrac{1}{2} T_{[AB|}{}^D F_{D|C)}{}^I \equiv 0 ~, \tag{6.3a}$$

$$\tfrac{1}{2}\nabla_{[A} G_{BC)}{}^i - \tfrac{1}{2} T_{[AB|}{}^D G_{D|C)}{}^i \equiv 0 ~, \tag{6.3b}$$

$$\tfrac{1}{2}\nabla_{[A} T_{BC)}{}^D - \tfrac{1}{2} T_{[AB|}{}^E T_{E|C)}{}^D \equiv 0 ~. \tag{6.3c}$$

Our superspace constraints at engineering dimensions $d \leq 1$ are

$$T_{\alpha\beta}{}^c = +2(\gamma^c)_{\alpha\beta} ~, \quad T_{\alpha\beta}{}^\gamma = T_{\alpha b}{}^c = T_{ab}{}^c = T_{ab}{}^\gamma = 0 ~, \tag{6.4a}$$

$$F_{\alpha b}{}^I = -(\gamma_b \lambda^I)_\alpha \ , \qquad F_{\alpha\beta}{}^I = 0 \ , \tag{6.4b}$$

$$\nabla_\alpha \lambda_\beta{}^I = +\tfrac{1}{2}(\gamma^{cd})_{\alpha\beta} F_{cd}{}^I + i(\gamma_5)_{\alpha\beta} D^I \ , \qquad \nabla_\alpha D^I = -i(\gamma_5 \slashed{\nabla} \lambda^I)_\alpha \ , \tag{6.4c}$$

$$G_{\alpha b}{}^i = -(\gamma_b \chi^i)_\alpha \ , \qquad G_{\alpha\beta}{}^i = 0 \ , \tag{6.4d}$$

$$\nabla_\alpha \chi_\beta{}^i = +\tfrac{1}{2}(\gamma^{cd})_{\alpha\beta} G_{cd}{}^i + i(\gamma_5)_{\alpha\beta} H^i \ , \qquad \nabla_\alpha H^i = -i(\gamma_5 \slashed{\nabla} \chi^i)_\alpha \ . \tag{6.4e}$$

The BIs at $d \geq 3/2$ yield field equations consistent with our lagrangian $\mathcal{L}_{VM}$ in (2.2).

This superspace reformulation provides an independent reconfirmation of the total consistency of our system. In particular, the validity of our field strengths (2.3) has been reconfirmed in superspace in (6.2), with two VMs gauging both generators $T^I$ and $U^i$

## 7. Summary and Concluding Remarks

In this paper, we have presented a new formulation for $N=1$ supersymmetric VMs in non-adjoint real representations of $SO(N)$. Our basic algebra is summarized into the commutators (2.1), and the condition for the possible real representation $R$ is (2.10). Accordingly, we have the new gauge field $B_\mu{}^i$ for the new generators $U^i$ in the representation $R$ of $SO(N)$. The system is consistent with supersymmetry under (2.10), and there seems to be no fundamental problem dealing with such a 'non-adjoint' vector field, in contrast to common wisdom.

The trivial example of $R$ is the **N** of $SO(N)$. However, there can be other *real* representations satisfying the condition (2.10), such as the spinorial $\mathbf{8}_S$ and conjugate spinorial $\mathbf{8}_C$ representations of $SO(8)$. Even though the $\mathbf{8}_V$, $\mathbf{8}_S$ and $\mathbf{8}_C$ of $SO(8)$ are naturally related by the triality, our formulation is the first one dealing with a VM in the spinorial representation of a gauge group. We emphasize that our system is *not* a rewriting of the usual supersymmetric $SO(N)$ gauge theory, when $R$ is such a non-trivial representation as the $\mathbf{8}_S$ of $SO(8)$. We have extra symmetry with $U^i$ accompanied by its proper gauge field $B_\mu{}^i$ with new freedom. To our knowledge, our system is the first one that has vector multiplets in non-adjoint representations with highly non-trivial interactions.

We have further coupled the system to CMs, and see that a Higgs mechanism can generate a mass to the new gauge field $B_\mu{}^i$. We have found that there are actually non-trivial v.e.v.'s that break $SO(N)$, while maintaining supersymmetry. We have also confirmed that the (mass)$^2$-matrices for all the vector fields have only positive definite eigenvalues, as desired. Interestingly, two different CMs $\Phi^I$ and $\Phi^i$ are needed for the total action to be $\delta_U$-invariant.

We have confirmed that the same formulation is possible in 10D, and thereby we can have $N=4$ supersymmetric theory with vectors in non-adjoint representations also in 4D. Remarkably, *maximally-extended* $N=4$ theory can be further coupled to extra multiplets in the non-adjoint representations. We have also performed superspace reformulation as an independent confirmation of the consistency of the whole idea. Our successful results here imply that there are more applications for our basic technique of treating non-adjoint VMs with supersymmetry.